\documentstyle[11pt] {article}
\newlength{\mytopmargin}
\newlength{\myleftmargin}
\begin{document}
\title{Exact ground state correlations in the supersymmetric\\
 $1/r^2$ exchange
$t-J$ model}

\author{P.J. Forrester\thanks{email: matpjf@maths.mu.oz.au}\\
Department of
Mathematics\\ University of Melbourne\\ Parkville\\ Victoria 3052\\ Australia}
\date{}
\maketitle
\begin{abstract} Some ground state static correlation functions for the $1/r^2$
exchange $t-J$ model of Kuramoto and Yokoyama are calculated exactly. The
correlations
calculated are the pair distribution functions between two down spins,
a down spin and a hole, and two holes, and the one body density matrix
(equal time retarded Green's function) for holes.\\
{}
\end{abstract}

It has now been established that the Calogero-Sutherland model provides an
exactly solvable example of one-dimensional anyon excitations [1,2]. This
has been achieved by the exact calculation of some ground state correlations,
which at rational coupling have been shown to involve only a finite number
of elementary excitations obeying anyon statistics [1-6]. Recently, Ha and
Haldane [7] have found that the supersymmetric $1/r^2$ exhange $t-J$
Hamiltonian of
Kuramoto and Yokoyama [8] shares this latter property. As a consequence, the
various static and dynamical ground state
correlations functions can be expressed as
integrals over rapidity variables in which the integrand has a compact support.
However the explicit form of the integrand could not be determined by the
numerical methods used in [7].

In this Letter we will provide explicit integral representations for the
down spin- down spin, down spin- hole and hole-hole two-point correlation
functions as well as the one-body density matrix for holes. The calculation
uses generalizations of integration techniques from the theory of random
matrices [9] which were developed and applied by the present author to the
study of binary plasma systems [10,11] more than ten years ago.

The $t-J$ Hamiltonian of Kuramoto and Yokoyama [8; see also 12,13] is
$$
{\cal H} = -\sum_{i<j,\sigma} t_{ij}c_{i\sigma}^\dagger c_{j\sigma}
+ \sum_{i<j} J_{ij} \left ( \vec{S_i}\cdot \vec{S_j} - {1 \over 4} n_i n_j
\right )
\eqno (1a)
$$
where
$$
t_{ij} = J_{ij} = \left ({\pi \over M} \right )^2 {1 \over \sin^2 \pi (i-j)/M
},
\eqno (1b)
$$
$M$ is the number of sites and double occupancy of the sites is implicitly
projected out. Let $x_1, \dots, x_{N_\downarrow}$ and $y_1,\dots,y_{N_\circ}$
specify
the location of the down spins ($\downarrow$) and holes ($\circ$) respectively,
where
$$
x_\alpha,y_j \in \{1,\dots,M\}
\eqno (2)
$$
Then the ground state wavefunction $\psi_0$ is given by [8]
\begin{eqnarray*}
\lefteqn{\psi_0(x_1,\dots,x_{N_\downarrow};y_1,\dots,y_{N_\circ})}\\
&=& \prod_{\alpha = 1}^{N_\downarrow} e^{- \pi i x_\alpha}
\prod_{j=1}^{N_\circ}
e^{- \pi i y_j} \prod_{1 \le \alpha < \beta \le N_\downarrow}
{\rm d}^2 (x_\alpha - x_\beta) \prod_{1 \le j < k \le N_\circ}
{\rm d}(y_k - y_j)
\end{eqnarray*}
$$
\times \prod_{\alpha = 1}^{N_\downarrow} \prod_{j=1}^{N_\circ} {\rm d}
(x_\alpha - y_j)
\eqno (3a)
$$
where
$$
{\rm d}(x) := 2 \sin \pi (x - x' )/M
\eqno (3b)
$$
This same wavefunction is also the exact ground state of the $1/r$ exchange
Hubbard model [14].

With $x_\alpha, y_j$ as continuous variables, we have previously
[11] interpreted $|\psi_0|^2$ as a generalised binary plasma  and calculated
the two-particle distribution functions exactly.
Use of the identity
$$
\int_0^M f(x/M) \, dx = \sum_{n=1}^M f(n/M)
\eqno (4a)
$$
whenever $f$ has the Fourier expansion
$$
f(x) = \sum_{k=-N}^N a_n e^{2 \pi i k x}, \qquad N < M,
\eqno (4b)
$$
shows that provided
$$
2 (N_{\downarrow} - 1 ) + N_\circ < M
\eqno (5)
$$
the ground state distributions of the continuous and discrete systems are the
same. Assuming (5), we can thus read off from [11]
that in the thermodynamic limit the static ground state two body correlations
of (1) are
$$
h_{\circ \circ} (r) = - | \int_0^1 dt \, e^{2 \pi i\rho_{\circ} r t}|^2
- {\rho_\downarrow \over \rho_\circ} \int_0^1 dt \int_0^1 ds \,
{1 - 2s \over 2 \rho_\downarrow t / \rho_\circ + 1}
\left ( e^{2 \pi i \rho_\circ r (s - 1 - \rho_\downarrow t / \rho_\circ)}
- e^{2 \pi i \rho_\circ r (s + \rho_\downarrow t /\rho_\circ ) } \right )
\eqno (6a)
$$
$$
h_{\downarrow \downarrow}(r)= \int_0^1 dt \int_0^1 ds \left [
- {(t + s + \rho_\circ/\rho_\downarrow )^2 \over (2t +
\rho_\circ/\rho_\downarrow )
(2s + \rho_\circ/\rho_\downarrow )} e^{ 2 \pi i\rho_\downarrow r(t-s)} \right .
\hspace{6cm}
$$
$$
\left. + {(t-s)^2 \over (2t + \rho_\circ/\rho_\downarrow )
(2s + \rho_\circ/\rho_\downarrow )} \cos \Big (2\pi \rho_\downarrow r
(t + s + \rho_\circ/\rho_\downarrow )\Big ) \right ]
\eqno (6b)
$$
$$
h_{\circ\downarrow}(r)= -2 \int_0^1 dt \int_0^1 ds \,
{t + \rho_\circ s/\rho_\downarrow \over 2t +  \rho_\circ/\rho_\downarrow }
\cos \Big (2 \pi \rho_\circ r (\rho_\downarrow t /\rho_\circ  - s +1 ) \Big)
\hspace{4cm}
\eqno (6c)
$$
where $r$ is an integer and
$$
\rho_\circ := N_\circ /M \qquad \rho_\downarrow = N_\downarrow/M
\eqno (6d)
$$

The double integrals (6) all have compact support as anticipated by Ha and
Haldane [7]. However, the forms of the integrands do not allow any
immediate interpretation in terms of the form factors for the elementary
excitations. This is also true of the expressions obtained from random
matrix theory for the static two-body correlation of the Calogero-Sutherland
model at $\lambda=2$ [15], which is the limiting case of (6b)
when $\rho_\downarrow = 0$. It wasn't until the discovery of a new integral
representation [3] that an interpretation in terms of the form
factors for the elementary excitations became possible [1]. However,
the formulas (6) are in a suitable form to study analytic properties of the
correlations. In particular, the asymptotic expansions of (6) show a
competition between
the two  periods $1/(\rho_\downarrow +\rho_\circ)$
and $1/\rho_\downarrow$ [11]. These periods are exhibited in the plots of Fig.
1.

The integrals (6) were obtained by taking the thermodynamic limit of the
finite system expressions. For example, in the finite system [16]
$$
h_{\downarrow \downarrow}(r)={1\over N_\downarrow(N_\downarrow-1)}
\sum_{p=1}^ {N_\downarrow} \sum_{q=1}^ {N_\downarrow} \left [
1 - {(2 N_\downarrow + N_\circ +1 -p -q)^2 e^{i 2 \pi (p - q) r/M}
\over (2 N_\downarrow + N_\circ +1 -2p)(2 N_\downarrow + N_\circ +1 -2q)}
\right .
$$
$$
\left .
+ {(p-q)^2 \cos 2 \pi(2 N_\downarrow + N_\circ +1 -p -q) \over
 (2 N_\downarrow + N_\circ +1 -2p)(2 N_\downarrow + N_\circ +1 -2q)}
\right ]
\eqno (7)
$$

It is also possible to calculate the density matrix (i.e. equal time retarded
Green's function) for the down spins, $-iG_{\downarrow \downarrow}(x-x')$
say. This quantity, having no classical analogue, wasn't considered in [11].
In a system with $N_\downarrow+1$ down spins, $N_\circ$ holes and the
inequality (5) obeyed,
$$
-iG_{\downarrow \downarrow}(x-x') =
C\prod_{\alpha = 1}^{N_\downarrow} \prod_{j=1}^{N_\circ}
\int_0^M dx_\alpha \int_0^M dy_j
\psi_0(x,x_1,\dots,x_{N_\downarrow};y_1,\dots,y_{N_\circ})
$$
$$
\times \psi_0(x',x_1,\dots,x_{N_\downarrow};y_1,\dots,y_{N_\circ})
\eqno (8)
$$
The normalization $C$ which is given explicitly in [11; see also 17], is chosen
so that $-iG_{\downarrow \downarrow}(0)= \rho_\downarrow$. It will be
convenient below to use the symbol $C$
 to denote the normalization in
this sense, even though its value may change due to the manipulations.

\setcounter{equation}{8}
{}From (8)
\begin{eqnarray}
\lefteqn{G_{\downarrow \downarrow}(x-x')} \nonumber\\
& & = C (-1)^{x - x'} e^{- \pi i (2 N_\downarrow + N_\circ)(x+x')/M}
\left ( \prod_{\alpha = 1}^{N_\downarrow}\int_0^M dx_\alpha
e^{- 2 \pi i x_\alpha (2 N_\downarrow + N_\circ)/M} \right ) \nonumber\\
& &  \times \left (\prod_{j=1}^{N_\circ}\int_0^M dy_j
e^{- 2 \pi i y_j ( N_\downarrow + N_\circ)/M} \right )
D(w,w',w_1,\dots,w_{N_\downarrow};z_1,\dots,z_{N_\circ})
\prod_{1 \le j < k \le N_\circ} (z_k - z_j)
\end{eqnarray}
where
\begin{eqnarray}
\lefteqn{D(w,w',w_1,\dots,w_{N_\downarrow};z_1,\dots,z_{N_\circ})}\nonumber\\
& & = \prod_{1 \le \alpha < \beta \le N_\downarrow} (w_\beta - w_\alpha)^4
\prod_{\alpha = 1}^{N_\downarrow} \prod_{j=1}^{N_\circ}(w_\alpha-z_j)^2
\prod_{1 \le j < k \le N_\circ}(z_k - z_j)^2\nonumber\\
& &\times \prod_{\alpha = 1}^{N_\downarrow} (w-w_\alpha)^2(w'-w_\alpha)^2
\prod_{j=1}^{N_\circ}(w-z_j)(w'-z_j)
\end{eqnarray}
with
$$
 w_\alpha = e^{2 \pi i x_\alpha /M} \qquad {\rm and} \qquad
z_j = e^{2 \pi i y_j /M}
$$
Next we recall from [11] that (10) can be expressed as the confluent
alternant determinant
\begin{equation}
{1 \over w - w'} \det \left [
\begin{array}{lllcl}
1&w_1&w_1^2&\cdots&w_1^{2N_\downarrow + N_\circ + 1} \\
0&1&2w_1&\cdots&(2N_\downarrow+N_\circ+1)w_1^{2N_\downarrow + N_\circ} \\
\vdots&\vdots&\vdots&\ddots&\vdots\\
1&w_{N_\downarrow}&w_{N_\downarrow}^2&\cdots&w_{N_\downarrow}^{2N_\downarrow +
N_\circ + 1} \\
0&1&2w_{N_\downarrow}&\cdots&(2N_\downarrow+N_\circ
+1)w_{N_\downarrow}^{2N_\downarrow
 + N_\circ}\\
1&w&w^2&\cdots&w^{2N_\downarrow+N_\circ+1}\\
1&w'&w'^2&\cdots&w'^{2N_\downarrow+N_\circ+1}\\
1&z_1&z_1^2&\cdots&z_1^{2N_\downarrow+N_\circ+1}\\
\vdots&\vdots&\vdots&\ddots&\vdots\\
1&z_{N_\circ}&z_{N_\circ}^2&\cdots&z_{N_\circ}^{2N_\downarrow+N_\circ+1}\end{array}
\right ]
\end{equation}
and that by the Vandermonde expansion
\begin{equation}
\prod_{1 \le j < k \le N_\circ}(z_k - z_j)=
\sum_{j=1}^{N_\circ !} \epsilon(P) \prod_{j=1}^{N_\circ}z_j^{P(j)-1}
\end{equation}

Since both (11) and (12) are antisymmetric in $\{z_1,\dots,z_{N_\circ}\}$ we
can replace
the l.h.s. of (12) in (9) by
\begin{equation}
\prod_{j=1}^{N_\circ}z_j^{j-1}
\end{equation}
which is the diagonal term on the r.h.s. of (12). The integrations over $z_j$
can
now be done row-by-row in the determinant (11). The columns $N_\downarrow +2$
to
$N_\downarrow + N_\circ+1$ inclusive of the final $N_\circ$ rows are then
diagonal.
Expansion by these columns, use of the definition of the remaining
determinant, and manipulation as in [11] gives
\begin{eqnarray}
\lefteqn{-iG_{\downarrow \downarrow}(x-x')}\nonumber \\
& & = C(-1)^{x - x'} e^{- \pi i (2 N_\downarrow + N_\circ)(x+x')/M}
\sum_{P=1 \atop P(2\alpha)>P(2\alpha-1)}^{(2 N_\downarrow + 2)!} \epsilon (P)
e^{2 \pi i (P(2 N_\downarrow + 1) - 1)x/M}
e^{2 \pi i (P(2 N_\downarrow + 2) - 1)x'/M} \nonumber \\
& & \times \prod_{\alpha=1}^{N_\downarrow} (P(2\alpha)-P(2\alpha-1))
\int_0^M dx_\alpha \, e^{2 \pi i x_\alpha (P(2 \alpha) + P(2 \alpha - 1) -
2N_\downarrow - N_\circ - 3)/M}
\end{eqnarray}

For non-zero contribution to the sum over permutations in (14) we require
\begin{equation}
P(2 \alpha) + P(2 \alpha - 1) = 2N_\downarrow + N_\circ + 3
\end{equation}
for $\alpha = 1, \dots, N_\downarrow$. Furthermore, since $P(2 \alpha)
> P(2 \alpha - 1)$, we must have
\begin{equation}
P(2 \alpha -1) \in \{1, \dots, N_\downarrow + 1\} \quad {\rm and } \quad
P(2\alpha) \in \{N_\downarrow + N_\circ + 2, \dots,2N_\downarrow + N_\circ +
2\}
\end{equation}
Setting
\begin{equation}
P(2 N_\downarrow + 1)=p \quad {\rm and} \quad
P(2 N_\downarrow + 2)=2N_\downarrow + N_\circ + 3-p, \quad
p \in \{1, \dots, N_\downarrow + 1\}
\end{equation}
we see that
\begin{equation}
\prod_{\alpha=1}^{N_\downarrow} (P(2\alpha)-P(2\alpha-1))=
{ \prod_{j=1}^{N_\downarrow +1}(2N_\downarrow + N_\circ + 3-2j) \over
2N_\downarrow + N_\circ + 3-2p}
\end{equation}
and is thus independent of the particular permutation. Hence, after minor
manipulation, (14) gives
\begin{equation}
-iG_{\downarrow \downarrow}(x-x') =
{C(-1)^{x-x'}\over \sin 2 \pi (x-x')/M} \sum_{p=1}^{N_\downarrow +1}{\sin \Big
(
\pi(2N_\downarrow + N_\circ + 3-p)(x-x')/M \Big) \over 2N_\downarrow + N_\circ
+ 3-2p}
\end{equation}
In the thermodynamic limit this becomes
\begin{equation}
-iG_{\downarrow \downarrow}(x-x') =
{(-1)^{x-x'} \over \pi (x-x')} \int_0^1
{\sin \Big( \pi(2 \rho_\downarrow t + \rho_\circ)(x-x')\Big ) \over
\rho_\circ/\rho_\downarrow + 2t} dt
\end{equation}
where $C$ has been chosen to give the correct normalization.

With $\rho_\circ = 0$ (20), apart from the sign $(-1)^{x - x'}$,
 reduces to the known expression [15] for the
equal time Green's function of the Calogero-Sutherland model at
$\lambda = 2$. A double integral representation is also known [3]. The
resulting integral identity allows us to express (20) in terms of double
integrals:
\begin{equation}
-iG_{\downarrow \downarrow}(x-x') = (-1)^{x - x'} \left [
( \rho_\downarrow  + \rho_\circ/2)I\Big((\rho_\downarrow  +
 \rho_\circ/2\Big )(x-x'))-
(\rho_\circ/2)I\Big (\rho_\circ(x-x')/2\Big ) \right ]
\end{equation}
where
\begin{equation}
I(x) := {1 \over 8} \int_{[-1,1]^2} dv_1 dv_2
{|v_1 - v_2| \over (1 - v_1^2)^{1/2}(1 -v_2^2)^{1/2}}
e^{i \pi x (v_1 + v_2)}
\end{equation}
Since (21) is the difference of two double integrals, it is not immediately
clear
how the integrand in (22) is related to the form factor.

In summary, we have calculated some static ground state correlations for the
$1/r^2$ exchange $t-J$ model. Our final expressions, as given by (6) and
(20), are  in terms of the double integrals over regions with compact
support. A complete interpretation in terms of the elementary excitations
remains, although the feature of the double integrals having compact
support was anticipated by Ha and Haldane [7].

\pagebreak

\noindent
{\bf References}

\begin{description}
\item[][1] F.D.M. Haldane, in {\it Proceedings of the 16th Taniguchi
Symposium} eds. A. Okiji and N. Kawakami (Springer-Verlag, 1994)
\item[][2] Z.N.C. Ha, cond-mat/9405063 submitted Phys. Rev. Lett.
\item[][3] P.J. Forrester, Phys. Lett. A {\bf 179} (1993) 127; cond-mat/
9408008 to appear J. Math. Phys.; cond-mat/ 9408042 to appear
Int. J. of Modern Physics B
\item[][4] J.A. Minahan and A.P. Polychronakos, hep-th/9404192 submitted
Phys. Lett. B
\item[][5] F. Lesage, V. Pasquier and D. Serban, hep-th/9405008 submitted
Nucl. Phys. B
\item[][6] E.R. Mucciolo, B.S. Shastry, B.D. Simons and B.L. Altshuler,
Phys. Rev. B {\bf 49} 15197 (1994)
\item[][7] Z.N.C. Ha and F.D.M. Haldane, cond-mat/9406059
\item[][8] Y.Kuramoto and H. Yokoyama, Phys. Rev. Lett. {\bf 67}, 1338 (1991)
\item[][9] M.L. Mehta, {\it Random Matrices}, 2nd ed. (Academic Press,
San Diego, 1991)
\item[][10] P.J. Forrester, J. Austr. Math. Soc. Series B {\bf 26}, 119 (1984)
\item[][11] P.J. Forrester and B. Jancovici, J. Physique Lett. {\bf 45},
{\bf L583} (1984)
\item[][12] D.F. Wang, J.T. Liu and P. Coleman, Phys. Rev. B {\bf 46}, 6639
(1992)
\item[][13] N. Kawakami, Phys. Rev. B {\bf 47}, 2928 (1993)
\item[][14] B. Sutherland, Phys. Rev. A, {\bf 2019} (1971)
\item[][15] D.F. Wang, Q.F. Zhong, P. Coleman, Phys. Rev. B {\bf 48}, 8476
(1993)
\item[][16] P.J. Forrester, Ph.D. thesis, Australian National University (1985)
\item[][17] D.M. Bressoud and I.P. Goulden, Commun. Math. Phys. {\bf 110}, 287
(1987)
\end{description}

\end{document}